\documentclass[a4paper,11pt]{article} 

\usepackage[english]{babel}
\usepackage{enumerate}  
\usepackage{amsmath}
\usepackage{pifont}
\usepackage{dsfont}
\usepackage{graphicx}
\usepackage{amsthm}
\usepackage{amssymb}

\usepackage{hyperref}    

\usepackage{ulem}  

\let\a=\alpha               
             \let\l=\lambda
                          \let\r=\rho

        \let\L=\Lambda

\def\cE{{\cal E}}

  \def\v0{{\vec 0}}

\def\bal{{\bar \l}}

\def\bR{\mathbb{R}}

\def\cE{\mathcal{E}}

\def\bN{\mathbb{N}}  
 
\def\bR{\mathbb{R}}

\def\indic{\hbox{\raise-2pt \hbox{\indbf 1}}}

\let\==\equiv

\let\0=\noindent

\def\*{{\hfill\break\null\hfill\break}}

\def\tende#1{\,\vtop{\ialign{##\crcr\rightarrowfill\crcr
			\noalign{\kern-1pt\nointerlineskip}
			\hskip3.pt${\scriptstyle #1}$\hskip3.pt\crcr}}\,}
\def\otto{\,{\kern-1.truept\leftarrow\kern-5.truept\to\kern-1.truept}\,}

\def\wt#1{\widetilde{#1}}

\def\sqt[#1]#2{\root #1\of {#2}}

\usepackage{tikz}
\usetikzlibrary{fit}
\usetikzlibrary{shapes.geometric}
\usetikzlibrary{decorations.pathmorphing}
\usetikzlibrary{decorations.pathreplacing,calligraphy}

\tikzset{
point/.style={circle,fill=black,inner sep=1pt},
vertex/.style={circle,fill=black,inner sep=1.5pt},   
bvertex/.style={circle,fill=black,inner sep=2.8pt},
Bvertex/.style={circle,fill=black,inner sep=4pt}, 
specialEP/.style={rectangle,fill=white,draw,inner sep=3pt},  
whitevex/.style={circle,fill=white,draw, inner sep=2pt},
linelabel/.style={sloped,above,very near start, inner sep=1pt,execute at begin node=$\scriptstyle,execute at end node=$},
baseline=(current  bounding  box.center),doubled/.style={double distance= 1pt,line width=1.5pt},
th/.style={line width=0.5 pt, gray},  
med/.style={line width=1 pt}  
}

\xdefinecolor{myred}{rgb}{0.7,0,0.2} 
\xdefinecolor{mypurple}{rgb}{0.6,0.2,0.4} 
\xdefinecolor{myblue}{rgb}{0.259,0.2,0.6}   
\xdefinecolor{myorange}{rgb}{1,0.6,0.2}   
\definecolor{orange}{rgb}{1,0.5,0}
\xdefinecolor{purpleish}{cmyk}{0.75,0.75,0,0}

\def\wt{\widetilde}

\def\be{\begin{equation}}
	\def\ee{\end{equation}}
\def\bea{\begin{eqnarray}}\def\eea{\end{eqnarray}}
\def\bean{\begin{eqnarray*}}\def\eean{\end{eqnarray*}}
\def\bfr{\begin{flushright}}\def\efr{\end{flushright}}
\def\bc{\begin{center}}\def\ec{\end{center}}
\def\bal{\begin{align}} 
	\def\eal{\end{align}}

\def\spl#1\spl{\[ \begin{split}#1\end{split} \]}

\def\bd{\begin{description}}\def\ed{\end{description}}
\def\bv{\begin{verbatim}}\def\ev{\end{verbatim}}

\def\Halmos{\hfill\vrule height10pt width4pt depth2pt \par\hbox to \hsize{}}

\newtheorem{theorem}{Theorem}[section]
\newtheorem{prop}{Proposition}[section]
\newtheorem{lemma}[prop]{Lemma} 
\theoremstyle{remark}

\numberwithin{equation}{section}
\def \aa{{\mathfrak a}}

\date{\today}

\title{Upper bound for the ground state energy \\ of a dilute Bose gas of hard spheres}

\author{Giulia Basti\footnote{Gran Sasso Science Institute, Viale Francesco Crispi 7, 67100 L'Aquila, Italy}\;, Serena Cenatiempo$^\ast$, Alessandro Giuliani\footnote{Universit{\`a} degli Studi Roma Tre, L.go S. Leonardo Murialdo~1, 00146 Roma, Italy}\;, Alessandro Olgiati\footnote{Institute of Mathematics, University of Zurich, Winterthurerstrasse 190, 8057 Zurich.}, \\ Giulio Pasqualetti$^\ddagger$, Benjamin Schlein$^\ddagger$}

\begin{document}

\maketitle

\abstract{We consider a gas of bosons interacting through a hard-sphere potential with radius $\frak{a}$ in the thermodynamic limit. We derive a simple upper bound for the ground state energy per particle at low density. Our bound captures the leading term $4\pi \rho \frak{a}$ and shows that corrections are smaller than $C \rho \aa (\rho \aa^3)^{1/2}$, for a sufficiently large constant $C > 0$. In combination with a known lower bound, our result implies that the first sub-leading term to the ground state energy is, in fact, of the order $\rho \aa (\rho \aa^3)^{1/2}$, in agreement with the Lee-Huang-Yang prediction.
}

\section{Introduction and Main Result} 

In the last years, there has been substantial progress in the mathematical understanding of the low-energy properties of dilute Bose gases. In the Gross-Pitaevskii regime, in which $N$ particles on the unit torus interact through a repulsive potential with range and scattering length of the order $1/N$, the ground state energy and the low-energy excitation spectrum have been determined in \cite{BBCS-acta}, up to errors vanishing as $N \to \infty$, under the assumption that the interaction potential $V \in L^3 (\bR^3)$ is repulsive, radial and of compact support. The proof applies optimal estimates on the number and the energy of excitations of the Bose-Einstein condensate that have been previously established in \cite{BBCS1,BBCS3}. Recently, a new derivation of these precise bounds has been proposed in \cite{HST}. The results of \cite{BBCS-acta} have been extended to systems of bosons trapped by an external potential (again in the Gross-Pitaevskii regime) in \cite{NNRT,BSS1,NT,BSS2}. They have been also generalized to the two-dimensional setting in \cite{CCS1,CCS2}. An upper bound on the ground state energy has been shown in \cite{BCOPS} for particles in the Gross-Pitaevskii regime, interacting through a hard-sphere potential. These results extend leading order estimates on the ground state energy that have been known since \cite{LSY,LSY2} and previous proofs of Bose-Einstein condensation obtained in  \cite{LS,LS2,NRS}. 

In the thermodynamic limit, where $N$ particles interacting through a repulsive potential $V$ with scattering length $\frak{a}$ are confined on a torus $\Lambda$ and $N, |\Lambda| \to \infty$ with fixed density $\rho = N / |\Lambda|$, the ground state energy per particle has been predicted by Lee-Huang-Yang in \cite{LHY} to satisfy 
\begin{equation}\label{eq:LHY} \lim_{\substack{N, |\Lambda| \to \infty: \\ \rho = N/|\Lambda|}} \frac{E_N}{N} = 4 \pi \frak{a} \rho \Big[ 1 + \frac{128}{15 \sqrt{\pi}} (\rho \frak{a}^3)^{1/2} + ... \Big] \end{equation} 
in the dilute regime $\rho \frak{a}^3 \to 0$. The validity of the leading order term on the r.h.s. of (\ref{eq:LHY}) was proven in \cite{Dy} (upper bound) and \cite{LY} (lower bound). An upper bound matching (\ref{eq:LHY}) was later shown  in \cite{YY} for sufficiently regular interaction potentials (improving an estimate previously shown in \cite{ESY}). Recently, a simpler proof applying to every repulsive and radial $V \in L^3 (\bR^3)$ was obtained in \cite{BCS}. A lower bound to the ground state energy matching (\ref{eq:LHY}) was established in \cite{FS1} for integrable potential and then in \cite{FS2}, also for hard-sphere interactions. In \cite{CJL, SimpleEq2}, the Lee-Huang-Yang formula (\ref{eq:LHY}) is proven, following a strategy proposed in \cite{L63}, under the assumption that the reduced densities associated with the ground state wave function satisfy certain relations. Although these relations have not yet been rigorously verified, they appear to capture the behaviour of Bose gases also beyond the dilute regime. Recently, a second order expansion for the ground state energy per particle of two dimensional Bose gases has been proven in \cite{FGJMO} for all positive potentials with finite scattering length.
The asymptotics of the ground state energy of dilute Fermi gases
was first studied in \cite{LSS}; for recent progress see \cite{FGHP, G, La, LaS}.

The derivation of an upper bound for the ground state energy resolving the Lee-Huang-Yang corrections in (\ref{eq:LHY}) for hard-sphere potentials remains open. In the present work, we make a step in this direction, providing a simple proof of the fact that the ground state energy per particle for hard-spheres in the thermodynamic limit is given by the leading term on the r.h.s. of (\ref{eq:LHY}), up to errors that are bounded above by 
$C \rho \aa (\rho \aa^3)^{1/2}$, in the limit $\rho \to 0$. Our result improves the upper bound obtained in \cite{Dy}, where the error was of the order $\rho \aa (\rho \aa^3)^{1/3}$.

We consider $N$ hard spheres moving in the box $\Lambda = [ - L/2 ; L/2 ]^3$, with periodic boundary conditions. We are interested in the limit $N,L \to \infty$ at fixed density $\rho = N / |\L|$. We define the ground state energy by 
\[ E^\text{hs}_N = \inf_{\Psi_N} \frac{\langle \Psi_N , \sum_{j=1}^N -\Delta_{x_j} \Psi_N \rangle}{\| \Psi_N \|^2} \]
where the infimum is taken over all $\Psi_N \in L^2_s (\Lambda^N)$, the subspace of $L^2 (\Lambda^N)$ consisting of functions that are symmetric w.r.t. permutations of the $N$ particles, satisfying the hard-sphere condition
\begin{equation}\label{eq:hard} \Psi_N (x_1, \dots , x_N) = 0 \qquad \text{if there exist $i, j \in \{ 1, \dots , N \}$ with } |x_i - x_j| \leq \frak{a}\,. \end{equation} 
Here $|x_i - x_j|$ denotes the distance on the torus between $x_i$ and $x_j$. 

\begin{theorem}\label{thm:main}  
There exists $C > 0$ such that 
\[ \lim_{\substack{N,L \to \infty : \\ N/ |\L| = \rho}} \frac{E^\text{hs}_N}{N} \leq 4\pi \rho \frak{a} \Big[ 1 + C (\rho \frak{a}^3)^{1/2} \Big] \]
for all $\rho \frak{a}^3 > 0$ small enough. 
\end{theorem}  

\section{The Trial State} 

In order to show Theorem \ref{thm:main}, we consider a wave function having the form 
\begin{equation}\label{eq:jastrow} \Psi_N (x_1, \dots , x_N) = \prod_{i<j}^N f_\ell (x_i - x_j)\,. \end{equation}
Such trial states have been first used in \cite{B,D,J}; for this reason we will refer to the product on the r.h.s. of (\ref{eq:jastrow}) as a Bijl-Dingle-Jastrow factor. In (\ref{eq:jastrow}), $f_\ell$ is chosen to describe correlations between particles, up to a distance $\ell \ll L$. More precisely, we choose $f_\ell$ to be the ground state solution of the Neumann problem
\begin{equation*}\label{eq:neumann} 
\left\{ \begin{split} 
-\Delta f_\ell &= \lambda_\ell f_\ell \\ 
f_\ell (x) &= 0 \qquad \text{for all } |x| < \frak{a} \\ 
\partial_r f_\ell (x) &= 0 \qquad \text{if } |x| = \ell 
\end{split} \right.  \end{equation*}
on the ball $B_\ell = \{ x \in \bR^3 :  |x| \leq \ell \}$, associated with the smallest eigenvalue $\lambda_\ell$. We normalize $f_\ell$ by requiring that $f_\ell (x) = 1$ for $|x| = \ell$. We extend $f_\ell$ to $\Lambda$, setting $f_\ell (x) =1$ for all $|x| \geq \ell$. We have then
\begin{equation}\label{eq:fell}  -\Delta f_\ell (x) = \lambda _\ell \chi_\ell (x) f_\ell (x) \end{equation} 
where $\chi_\ell$ denotes the characteristic function of the ball $B_\ell$. The proof of the following lemma can be found  in \cite[Lemma 2.1]{BCOPS} (it is easy to translate the bounds on $\omega_\ell = 1 - f_\ell$ stated in \cite{BCOPS} into the estimates for $u = 1- f^2_\ell$ appearing here).  
\begin{lemma} \label{lm:lambdaell}
For $\frak{a} \ll \ell$, we have 
 \begin{equation*} \label{eq:lambdaell-exp}
\lambda_\ell =\frac{3 \mathfrak{a}}{\ell^3} \Big[ 1+\mathcal{O} (\frak{a} / \ell ) \Big] 
\end{equation*}
Moreover, $0 \leq f_\ell (x) \leq 1$ for all $x \in \Lambda$ and, defining $u (x) = 1 - f^2_\ell (x)$, we find 
\[ 0 \leq u (x) \leq C \frak{a}  \frac{\chi_\ell (x)}{|x|} , \qquad |\nabla u (x) | \leq C \frak{a}  \frac{\chi_\ell (x)}{|x|^2}\,. \]
 \end{lemma}

Since (\ref{eq:jastrow}) satisfies the hard-core condition (\ref{eq:hard}), we immediately obtain that 
\[ E_N^\text{hs} \leq \sum_{j=1}^N  \frac{\langle \Psi_N, -\Delta_{x_j} \Psi_N \rangle}{\| \Psi_N \|^2} \,. \]
For $j=1, \dots , N$, we compute 
\[ \begin{split}  -\Delta_{x_j} \Psi_N (x_1, \dots , x_N) = \; &\sum_{i \not = j}^N \frac{-\Delta f_\ell (x_j - x_i)}{f_\ell (x_j - x_i)}  \Psi_N (x_1, \dots , x_N) \\ &- \sum^N_{\substack{i,m \not = j \\ i \not = m}} \frac{\nabla f_\ell (x_j - x_i)}{f_\ell (x_j -x_i)} \cdot \frac{\nabla f_\ell (x_j - x_m)}{f_\ell (x_j - x_m)}  \Psi_N (x_1, \dots , x_N)\,.
\end{split} \]
From (\ref{eq:fell}), we obtain 
\[ \begin{split} \langle \Psi_N, -\Delta_{x_j} \Psi_N \rangle =\; & \sum_{i \not = j}^N \int  \lambda_\ell \chi_\ell (x_j- x_i) |\Psi_N (x_1 \dots , x_N)|^2 \, dx_1 \dots dx_N \\ &- \sum_{\substack{i,m \not = j \\ i \not = m}} \int \frac{\nabla f_\ell (x_j - x_i)}{f_\ell (x_j -x_i)} \cdot \frac{\nabla f_\ell (x_j - x_m)}{f_\ell (x_j - x_m)}  |\Psi_N (x_1, \dots , x_N)|^2\, dx_1 \dots dx_N \,. \end{split} \]
For $i,j \in \{ 1, \dots , N \}$, we write $V_{ij} = 2 \lambda_\ell \chi_\ell (x_i - x_j)$ and $f_{ij} = f_\ell (x_i - x_j)$. With this short-hand notation (and omitting the measure $dx_1 \dots dx_N$ from all integrals), we find 
\[ \frac{E_N^\text{hs}}{N} \leq \frac{(N-1)}{2} \frac{\int V_{12} \prod_{i<j}^N f_{ij}^2}{\int \prod_{i<j}^N f_{ij}^2}  - \frac{(N-1)(N-2)}{6} \frac{\int \frac{\nabla f_{13}}{f_{13}} \cdot \frac{\nabla f_{23}}{f_{23}} \prod_{i<j}^N f_{ij}^2}{\int \prod_{i<j}^N f_{ij}^2}\,.  \]
The two terms on the r.h.s. of the last equation will be considered in the next two propositions, whose proof is deferred to the next sections.

\begin{prop} \label{prop:2body}
Fix $\ell = c \, (\r \aa)^{-1/2}$ for a sufficiently small constant $c > 0$. Then there is a constant $C > 0$ such that 
\begin{equation}\label{eq:prop1}
\limsup_{\substack{N,|\Lambda| \to \infty : \\ N/|\Lambda| = \rho}} \frac{N}{2} \frac{\int V_{12} \prod_{i<j}^N f_{ij}^2}{\int \prod_{i<j}^N f_{ij}^2} \leq 4 \pi \aa \r    + C  \r \aa (\r \aa^3)^{1/2}
\end{equation}
for all $\rho \frak{a}^3 > 0$ small enough. 
\end{prop}

\begin{prop}\label{prop:3body}
Fix $\ell = c \, (\r \aa)^{-1/2}$ for a sufficiently small constant $c > 0$. Then there is a constant $C > 0$ such that 
\[
\limsup_{\substack{N,|\Lambda| \to \infty : \\ N / |\Lambda| = \rho}} \left| N^2 \frac{\int \frac{\nabla f_{13}}{f_{13}} \cdot \frac{\nabla f_{23}}{f_{23}} \prod_{i<j}^N f_{ij}^2}{\int \prod_{i<j}^N f_{ij}^2}\right| \leq  C \r \aa (\r \aa^3)^{1/2}\
\] 
for all $\rho \frak{a}^3 > 0$ small enough. 
\end{prop}

From Prop. \ref{prop:2body} and Prop. \ref{prop:3body} (and from the existence of the thermodynamic limit for the energy per particle $E_N^\text{hs}/N$), we immediately conclude that there exists $C > 0$ such that 
\[ 
\lim_{\substack{N,|\Lambda| \to \infty: \\ N / |\Lambda| = \rho}} \frac{E_N^\text{hs}}{N} \leq 4 \pi \aa \r  \left[ 1 + C (\r \aa^3)^{1/2} \right] \]
for all $\rho \frak{a}^3 > 0$ small enough. This completes the proof of Theorem \ref{thm:main}. 

\medskip

The proof of Prop. \ref{prop:2body} and Prop. \ref{prop:3body} is based on rewriting $f^2_{ij}=1-u_{i,j}$ in the Bijl-Dingle-Jastrow factor, on expanding it in powers of $u_{i,j}$, and on identifying precise cancellations between the numerator and the denominator. The various terms in the expansion can be graphically represented as diagrams in which nodes represent particles' labels and lines connecting nodes correspond to factors $u_{i,j}$. There are two kinds of cancellations between the diagrams at the numerator and at the denominator. One is standard, and is at the very root of the cluster expansion method: all disconnected diagrams cancel between numerator and denominator, and one is left with an expansion over connected diagrams only. This cancellation is not enough for proving that the error term in \eqref{eq:prop1}
is of relative order $(\rho \aa^3)^{1/2}$, but `just' $(\rho\aa^3)^{1/3}$, the same as the error term in Dyson's upper bound \cite{Dy}. In order to go beyond this one needs to identify additional, more subtle, cancellations. Explicit computations at low orders show that all tree diagrams cancel between numerator and denominator: this suggests that only connected diagrams {\it with loops} should survive. Actually all `reducible' diagrams cancel at the first few orders, but the cancellations of trees is sufficient to obtain an error term comparable with the Lee-Huang-Yang correction. The cancellation of reducible diagrams was already noticed by Jastrow, see \cite[Eqs.\,(11)--(11c)]{J} and is explicitly discussed in  \cite[below Eq.\,(3.6)]{PB}, even though not proved systematically. Its rigorous proof has been obtained 
much more recently within a convergent cluster expansion scheme in the canonical ensemble \cite{PT}.
In this paper, instead of using a standard cluster expansion, we find it more convenient to perform partial expansions of numerator and denominator, choosing the order of the expansion large enough for the truncation errors to be small. At each step of the iteration, we estimate contributions associated with loop diagrams and we isolate fully expanded trees, whose contribution is going to cancel when we combine the estimates we obtain for the numerator and the denominator.

\section{Proof of Proposition \ref{prop:2body}}

We set $\ell = c \, (\rho \frak{a})^{-1/2}$ for a sufficiently small constant $c > 0$ to be specified later on. 
Then 
\begin{equation}\label{eq:ell} \rho \frak{a} \ell^2 = c^2 \ll 1 \, . \end{equation}  
We introduce the notation 
\begin{equation}\label{eq:INk}  I_{N-k} = \int \prod_{k+1 \leq  i  < j \leq N} f_{ij}^2 \, dx_{k+1} \dots dx_N  \end{equation} 
for $k =0,1, \dots , N-2$. We observe that $I_{N-k} \leq I_{N-(k+1)} |\Lambda|$ for all $k =0, \dots , N-3$. At the same time, defining $u (x) = 1- f_\ell^2 (x)$ and using Lemma \ref{lm:lambdaell} to estimate $\| u \|_1 \leq C \frak{a} \ell^2$, we find  
\[ I_{N-k} \geq I_{N-(k+1)} (|\Lambda| - C N \| u \|_1) \geq |\Lambda| I_{N- (k+1)} (1 - C \rho \frak{a} \ell^2) \geq |\Lambda | I_{N- (k+1)} / 2 \]
choosing $c > 0$ in (\ref{eq:ell}) small enough. Repeating the same argument, we obtain \begin{equation}\label{eq:INm} 2^{-m} |\Lambda|^m I_{N-(k+m)} \leq I_{N-k} \leq |\Lambda|^m I_{N-(k+m)} \end{equation} 
for all $k,m \in \bN$ with $k+m \leq N-2$. 

We consider the numerator on the l.h.s. of (\ref{eq:prop1}). We isolate the term $f_{12}^2$ and we expand the remaining $x_1$-dependence in the Bijl-Dingle-Jastrow factor, defining $u_{i,j} = u (x_i - x_j) = 1 - f^2_{ij}$, for any $1 \leq i < j \leq N$. Choosing $M \in \bN$ even to make sure that the last term in the expansion is positive (as needed to have an upper bound since $u \geq 0$), we obtain  
\begin{equation}\label{eq:step1} \begin{split} &\int V_{12}  \prod_{1 \leq i,j \leq N} f_{ij}^2  \\ &\leq \int V_{12} f_{12}^2 \Big[ 1 - \sum_{3 \leq r_1 \leq N} u_{1,r_1} +  \dots + \sum_{3 \leq r_1 < r_2 < \dots < r_{M} \leq N} u_{1,r_1} \dots u_{1, r_{M}} \Big] \prod_{2 \leq i < j \leq N} f_{ij}^2 
\\ &= \sum_{m_1 = 0}^{M} (-1)^{m_1} \sum_{3 \leq r_1 < r_2 < \dots < r_{m_1} \leq N} \int V_{12} f_{12}^2 u_{1,r_1} \dots u_{1, r_{m_1}}  \prod_{2 \leq i < j \leq N} f_{ij}^2 
\\ &= \sum_{m_1 = 0}^{M} (-1)^{m_1} {N-2 \choose m_1} \int V_{12} f_{12}^2 u_{1,3} u_{1,4} \dots u_{1,m_1+2} \prod_{2 \leq i < j \leq N} f_{ij}^2\,. \end{split} \end{equation} 
Here, and similarly below, we use the convention that, if $m_1 = 0$, there is no factor of $u$ in the integral. Next, we expand the $x_2$-dependence in the Bijl-Dingle-Jastrow factor. We find (here, we stop the expansion at $m_2 = M - m_1$, which again guarantees that the last term is positive) 
\begin{equation}\label{eq:step2} \begin{split} 
\int &V_{12} \prod_{1 \leq i,j \leq N} f_{ij}^2 \\ &\leq \sum_{m_1 = 0}^{M} (-1)^{m_1} {N-2 \choose m_1} \sum_{m_2 = 0}^{M - m_1} (-1)^{m_2} \\ &\hspace{1cm} \times  \sum_{3 \leq r_1 < \dots < r_{m_2} \leq N} \int V_{12} f_{12}^2 u_{1,3} \dots u_{1,m_1+2} \, u_{2,r_1} \dots u_{2, r_{m_2}} \prod_{3 \leq i < j \leq N} f_{ij}^2  \,.
\end{split} \end{equation} 

Furthermore, we get rid of the contribution of the loops, namely of the terms where there exists at least one index $i \in \{1, \ldots, m_2\}$ with $r_i \in \{3, \ldots, m_1+2$\}. We find
\begin{equation}\label{eq:V12-1} \begin{split} 
\int V_{12} &  \prod_{1 \leq i,j \leq N} f_{ij}^2 \\ \leq \; &\sum_{m_1 = 0}^{M} (-1)^{m_1} {N-2 \choose m_1} \sum_{m_2 = 0}^{M - m_1} (-1)^{m_2}  {N-2-m_1 \choose m_2} \\ &\hspace{1cm} \times  \int V_{12} f_{12}^2 u_{1,3} \dots u_{1,m_1+2} u_{2,m_1+3} \dots u_{2, m_1 + m_2 + 2} \prod_{3 \leq i < j \leq N} f_{ij}^2  \\ &+ \cE_{\text{loops},2} 
\end{split} \end{equation} 
where (denoting by $k$ the number of loops) 
\[ \begin{split}
\cE_{\text{loops},2} &\; = \sum_{m_1 = 1}^{M} (-1)^{m_1} {N-2 \choose m_1} \sum_{m_2 = 1}^{M - m_1} (-1)^{m_2}  \sum_{k=1}^{\min (m_1, m_2)} {m_1 \choose k}{N-2 - m_1 \choose  m_2 -k }  \\
&\hskip 1cm \times  \int V_{12} f^2_{12} u_{1,3} \ldots u_{1,k+2} u_{2,3} \ldots u_{2,k+2}\\ 
& \hskip 2cm  \times   u_{1, k+3} \ldots u_{1, m_1+2}  u_{2, m_1 +3} \ldots u_{2,  m_1+ m_2 +2-k} \prod_{3 \leq i <j \leq N} f^2_{ij}\,.
\end{split}\]
\begin{figure}[t] 
\centering
\begin{minipage}{10.2cm} \vskip -1cm
\begin{minipage}{5cm}
\begin{tikzpicture}[scale =0.7,
    every path/.style = {},
 ]
  \begin{scope} 
  \filldraw[black] (1.42,-0.45) circle (3pt); 
  \node[right] at (1.45, -0.3) {\footnotesize  $3$};
  \draw[ thick,  - ] (0,0) -- (1.42,-0.45); 
  \filldraw[black] (1.2,-0.9) circle (3pt); 
       \node[below] at (1.4, -0.8) {\footnotesize  $4$};
  \draw[ thick,  - ] (0,0) -- (1.2,-0.9);
  \draw[ thick, dashed, - ] (0,0) -- (0.7,-1.35);
   \filldraw[black] (0,-1.5) circle (3pt) node[below]{\footnotesize  $m_1+2$};
    \draw[ thick,  - ] (0,0) -- (0,-1.5);
    \filldraw[black] (3,0) circle (3pt); 
  \draw[ thick,  - ] (1.5,0) -- (3,0);
  \filldraw[black] (2.9,0.5) circle (3pt);
 \draw[ thick,  - ] (1.5,0) -- (2.9,0.5); 
  \draw[ thick,  -, dashed ] (1.5,0) -- (2.65,0.9);
   \filldraw[black] (2.3,1.25) circle (3pt);
  \draw[ thick,  - ] (1.5,0) -- (2.3,1.25);
  \draw [thick, decorate, decoration={brace,amplitude=5pt,raise=-0.7ex}]
  (2.8,1.65) -- (3.6,0.1) node[midway]{};
    \node[right] at (3.2, 1) {\footnotesize  $m_2$};
    \filldraw[black] (2.8,-1) circle (3pt); 
  \draw[ thick,  - ] (1.42,-0.45) -- (2.8,-1);
  \draw[ thick,  -, dashed ] (1.42,-0.45) -- (2.6,-1.45);
        \filldraw[black] (2.18,-1.72) circle (3pt); 
  \draw[ thick,  - ] (1.42,-0.45) -- (2.18,-1.72);
    \draw [thick, decorate, decoration={brace,amplitude=5pt,raise=1ex}]
  (2.9,-1) -- (2.2,-1.8) node[midway]{};
  \node[right] at (2.9, -1.9) {\footnotesize  $m_3$};
   \filldraw[red] (0,0) circle (3pt); 
    \node[above, color=red] at (0, 0.05) {\footnotesize  $1$};
  \filldraw[red] (1.5,0) circle (3pt); 
   \node[above, color=red] at (1.4, 0.05) {\footnotesize  $2$};
  \draw[ thick,  -, color=red, very thick, dotted ] (0,0) -- (1.5,0);
\end{scope}
\end{tikzpicture}
\end{minipage}
\begin{minipage}{5cm} \vskip 1.2cm
\begin{tikzpicture}[scale =0.75,
    every path/.style = {},
 ]
  \begin{scope} 
  \draw[very thick, color=lightgray] (0, -1.5) to [out=-20,in=-90] (1.2,-0.9);
    \draw[very thick, color=lightgray] (1.2, -0.9) to [out=-70,in=-45, distance=3.5cm] (2.9,0.5);
   \filldraw[lightgray] (1.45,-2.36) circle (3pt); 
   \draw[ very thick,  - , color=lightgray] (1.2,-0.9) --  (1.45,-2.36);
    \filldraw[lightgray] (0.95,-2.36) circle (3pt); 
   \draw[ very thick,  - , color=lightgray] (1.2,-0.9) --  (0.95,-2.36);

  \filldraw[black] (1.42,-0.45) circle (3pt); 
  \node[right] at (1.45, -0.3) {\footnotesize  $3$};
  \draw[ thick,  - ] (0,0) -- (1.42,-0.45); 
  \filldraw[black] (1.2,-0.9) circle (3pt); 
    \node[below] at (1.45, -0.75) {\footnotesize  $4$};
  \draw[ thick,  - ] (0,0) -- (1.2,-0.9);
  \draw[ thick, dashed, - ] (0,0) -- (0.7,-1.35);
   \filldraw[black] (0,-1.5) circle (3pt) node[below]{\footnotesize  $m_1+2$};
    \draw[ thick,  - ] (0,0) -- (0,-1.5);
    \filldraw[black] (3,0) circle (3pt); 
  \draw[ thick,  - ] (1.5,0) -- (3,0);
  \filldraw[black] (2.9,0.5) circle (3pt);
 \draw[ thick,  - ] (1.5,0) -- (2.9,0.5); 
  \draw[ thick,  -, dashed ] (1.5,0) -- (2.65,0.9);
   \filldraw[black] (2.3,1.25) circle (3pt);
  \draw[ thick,  - ] (1.5,0) -- (2.3,1.25);
    \filldraw[black] (2.8,-1) circle (3pt); 
  \draw[ thick,  - ] (1.42,-0.45) -- (2.8,-1);
  \draw[ thick,  -, dashed ] (1.42,-0.45) -- (2.6,-1.45);
        \filldraw[black] (2.18,-1.72) circle (3pt); 
  \draw[ thick,  - ] (1.42,-0.45) -- (2.18,-1.72);
   \filldraw[red] (0,0) circle (3pt); 
    \node[above, color=red] at (0, 0.05) {\footnotesize  $1$};
  \filldraw[red] (1.5,0) circle (3pt); 
   \node[above, color=red] at (1.4, 0.05) {\footnotesize  $2$};
  \draw[ thick,  -, color=red, very thick, dotted ] (0,0) -- (1.5,0);
\end{scope} 
\end{tikzpicture}
\end{minipage}
\end{minipage} \vskip -1cm
\begin{minipage}{14cm}
\caption{\small Graphical representation of the iterative expansion described between Eq.\,\eqref{eq:step1} and  Eq.\,\eqref{eq:2body-indu}. Nodes represent particles' labels, and a link between node $i$ and node $j$ represents a factor $u_{i,j}$, with the exception of the dotted link, which represents $V_{12}f^2_{12}$. 
On the l.h.s. an example of a diagram without loops obtained by expanding the $x_1, x_2$ and $x_3$-dependence in the Bijl-Dingle-Jastrow factor.  On the r.h.s. an example of a diagram with $k=2$ loops which is obtained from the previous one by expanding the $x_4$-dependence (the lines coming from the latter expansion are depicted in light grey). The two loops have lengths $s_1=3$ and $s_2=4$, respectively (in general, all loops have by construction length $s\ge 3$); the loop of length $4$ contains the dotted link,
while the other does not.} \label{Fig1}
\end{minipage}
\end{figure}
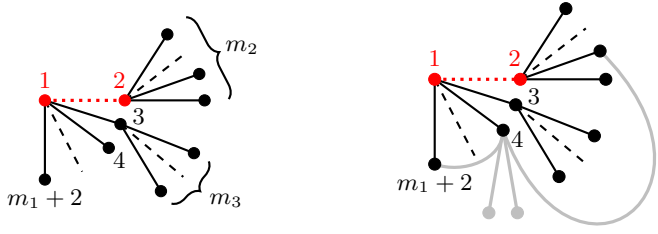

From Lemma \ref{lm:lambdaell}, we have $u (x) \leq C\frak{a} \chi_\ell (x) / |x|$. Thus, we can estimate 
\[ \begin{split}
& \Big| \int V_{12} f^2_{12} u_{1,3} \ldots u_{1,k+2} u_{2,3} \ldots u_{2,k+2} \, dx_1\cdots dx_{k+2} \Big|   \\
&  \leq C^{k} \aa^{2k}  \l_\ell |\Lambda| \int \chi(|x|\leq \ell)   \prod_{j=1}^k \frac{\chi(|y_j|\leq \ell)}{|y_j|}  \frac{\chi(|x+y_j|\leq \ell)}{|x+y_j|} \, dx dy_1 \dots dy_k \leq C \aa|\Lambda| (C \aa^2 \ell)^{k}\,.
\end{split}\]
for a constant $C > 0$ independent of all parameters. Using again the bound in Lemma \ref{lm:lambdaell} to show that $\| u \|_1 \leq C \frak{a} \ell^2$ and (\ref{eq:INm}), this implies that 
\[ \begin{split} N \cE_{\text{loops},2} &\leq C \frak{a} |\Lambda| \sum_{m_1 = 1}^M \frac{1}{m_1!} \sum_{m_2 = 1}^{M-m_1} \sum_{k=1}^{\min (m_1, m_2)}  \binom{m_1}{k}  \frac{1}{(m_2-k)!} \\ & \hspace{1cm} \times N^{m_1 + m_2 +1 -k} \| u \|_1^{m_1 + m_2 -2k} (C \frak{a}^2 \ell)^k I_{N-(m_1 + m_2 +2-k)} \\ &
\leq C \rho \frak{a} I_N \sum_{m_1=1}^M \frac{1}{m_1!}  \sum_{m_2= 1}^{M-m_1} \sum_{k=1}^{\min (m_1, m_2)}   \binom{m_1}{k} \frac{1}{(m_2-k)!}  (C \rho \frak{a} \ell^2)^{m_1 + m_2 - 2k} (C \rho \frak{a}^2 \ell)^k 
\end{split} \]
with an appropriate choice of the constant $C> 0$. Exchanging the sums over $k$ and $m_2$, and shifting $m_2 \to m_2 + k$, we arrive at 
\[ \begin{split} N \cE_{\text{loops},2} &\leq C \rho \frak{a} I_N \sum_{m_1=1}^M \sum_{k=1}^{m_1} \sum_{m_2 = 0}^{M-m_1-k} \binom{m_1}{k} \frac{1}{m_2!}  (C \rho \frak{a} \ell^2)^{m_1 + m_2 - k} (C \rho \frak{a}^2 \ell)^k \\
 &\leq C \rho \frak{a} I_N \sum_{m_1=1}^M \sum_{k=1}^{m_1} \binom{m_1}{k}   (C \rho \frak{a} \ell^2)^{m_1 - k} (C \rho \frak{a}^2 \ell)^k \\
 &\leq C \rho \frak{a} I_N \sum_{m_1=1}^M \sum_{k=1}^{m_1} \binom{m_1}{k}   (C \rho \frak{a} \ell^2)^{m_1} (C \frak{a} / \ell)^k  \leq C \rho \frak{a} (\rho \frak{a}^2 \ell) I_N 
\end{split} \]

In (\ref{eq:V12-1}), we also separate terms with $m_1 + m_2 = 0$ (in this case, there is only the term with $m_1 = m_2 = 0$), where the Bijl-Dingle-Jastrow factor is no longer entangled with the observable, from the other terms. We obtain 
\begin{equation}\label{eq:V12-x3} \begin{split} 
\int V_{12} &  \prod_{1 \leq i,j \leq N} f_{ij}^2 \\ \leq \; & I_{N-2} \, |\Lambda | \Big[  2 \lambda_\ell \int \chi_\ell (x)  f_\ell^2 (x) dx \Big] \\ &+ \sum_{m_1 = 0}^{M} (-1)^{m_1} {N-2 \choose m_1} \sum_{m_2 = 0}^{M - m_1} (-1)^{m_2}  {N-2-m_1 \choose m_2}  \chi (m_1 + m_2 \geq 1) \\ &\hspace{1cm} \times \int V_{12} f_{12}^2 u_{1,3} \dots u_{1,m_1+2} u_{2,m_1+3} \dots u_{2, m_1 + m_2 + 2} \prod_{3 \leq i < j \leq N} f_{ij}^2  \\ &+  C \rho \frak{a} (\rho \frak{a}^2 \ell)  I_{N} / N \,.
\end{split} \end{equation} 

Proceeding by induction (see Fig.\ref{Fig1} for a graphical representation of our expansion of the Bijl-Dingle-Jastrow factor) we claim that, for every $h \in \bN$, $h \geq 2$,  
\begin{equation}\label{eq:2body-indu}
\begin{split} 
\int V_{12}   \prod_{1 \leq i<j \leq N} f_{ij}^2  \leq \; & |\Lambda| \Big[ 2\lambda_\ell \int \chi_\ell (x) f^2_\ell (x) dx \Big] \Big[ I_{N-2}  +  \sum_{k=3}^h  \a_k I_{N-k}  \| u \|_1^{k-2}  \Big] \\ &+ \int V_{12} f_{12}^2 \, \beta_h \prod_{h+1 \leq i < j \leq N} f_{ij}^2 + C \rho \frak{a}^2 \ell^{-1} \sum_{j=2}^h (C \rho \frak{a} \ell^2)^{j-2}  I_N / N  
\end{split} \end{equation} 
where we defined 
\begin{equation*}\label{eq:alk} \begin{split} \alpha_k = \; &\sum_{m_1=0}^M (-1)^{m_1} {N-2 \choose m_1} \dots \hspace{-.2cm}\sum_{m_{k-1} = 0}^{M-m_1-\dots - m_{k-2}} (-1)^{m_{k-1}} {N-2-m_1-\dots -m_{k-2} \choose m_{k-1}} \\ &\hspace{1cm} \times \Big[ \prod_{j=2}^{k-2} \chi (m_1 + \dots + m_j \geq j-1)  \Big] \, \chi (m_1 + \dots + m_{k-1} = k-2) \end{split} \end{equation*} 
and 
\[ \begin{split}  
\beta_h = &\sum_{m_1=0}^M (-1)^{m_1} {N-2 \choose m_1} \dots  \sum_{m_h = 0}^{M-m_1-\dots - m_{h-1}} (-1)^{m_h} {N-2-m_1-\dots -m_{h-1} \choose m_{h}} \\ &\hspace{.1cm} \times \Big[ \prod_{j=2}^h \chi (m_1 + \dots + m_j \geq j-1) \Big]  \prod_{j_1 =3}^{m_1+2} u_{1,j_1}  \prod_{j_2 =m_1+3}^{m_1+m_2+2} u_{2,j_2} \dots \prod_{j_h =m_1+\dots + m_{h-1} +3}^{m_1+ \dots +m_h +2} u_{h,j_h}\,. \end{split}\] 
Notice that, by definition, $\beta_h$ is a sum of functions depending at least on the variables $x_1, \dots , x_{h+1}$ (more precisely, the term with the indices $m_1, \dots , m_h$ depends on the variables $x_1, \dots , x_{m_1+\dots + m_h+2}$). 

The bound (\ref{eq:V12-x3}) shows the validity of (\ref{eq:2body-indu}) with $h=2$, since $\rho \frak{a}^2 \ell = \frak{a} \ell^{-1} (\rho \frak{a} \ell^2) \leq \frak{a} \ell^{-1}$. To show the induction step we start from the bound (\ref{eq:2body-indu}) and, in the term  proportional to $\beta_h$, we expand the dependence of the Bijl-Dingle-Jastrow factor on the $x_{h+1}$ variable, similarly as we did in (\ref{eq:step2}). We obtain 
\begin{equation}\label{eq:ind} \begin{split} 
\int V_{12} &f_{12}^2  \beta_h \prod_{h+1 \leq i < j \leq N} f_{ij}^2  \\ \leq\; &\sum_{m_1=0}^M (-1)^{m_1} \binom{N-2}{m_1} \dots \sum_{m_h=0}^{M-m_1-\dots-m_{h-1}} (-1)^{m_h} \binom{N-2-m_1-\dots-m_{h-1}}{m_h} \\
		&\quad \times \left[ \prod_{j=2}^h \chi(m_1+\dots+m_j \ge j-1) \right] \sum_{m_{h+1} = 0}^{M-m_1-\dots - m_h} (-1)^{m_{h+1}}  \sum_{h+2 \leq r_1 < \dots < r_{m_{h+1}} \leq N}  \\ &\quad \times \int V_{12} f_{12}^2 \prod_{j_1=3}^{m_1+2} u_{1,j_1} \dots \prod_{j_h = m_1+\dots +m_{h-1} + 3}^{m_1 + \dots + m_h +2} u_{h, j_h} \prod_{j=1}^{m_{h+1}} u_{h+1, r_j}  \, \prod_{h+2 \leq i < j \leq N} f_{ij}^2 \,.	\end{split}\end{equation} 

As we did above, we separate terms with no loops from terms with at least one loop. We decompose the contribution arising from terms without loops writing $1 = \chi (m_1 + \dots + m_{h+1} \geq h) + \chi (m_1 + \dots + m_{h+1} = h-1)$ (we can restrict our attention to the support of $\chi (m_1 + \dots + m_h \geq h-1)$). We conclude that 
\begin{equation}\label{eq:ind-1} \begin{split} 
\int V_{12} f_{12}^2  \beta_h \prod_{h+1 \leq i < j \leq N} f_{ij}^2  \leq \; &|\Lambda|  \Big[ 2\lambda_\ell \int \chi_\ell (x) f^2_\ell (x) dx \Big]   \a_{h+1} I_{N-(h+1)}  \| u \|_1^{h-1} \\ &+ \int V_{12} f_{12}^2 \, \beta_{h+1} \prod_{h+2 \leq i < j \leq N} f_{ij}^2 + \cE_{\text{loops},h+1} 
\end{split} \end{equation} 
where $\cE_{\text{loops},h+1}$ denotes the contribution from terms on the r.h.s. of (\ref{eq:ind}) with at least one loop. 
Consider, for fixed $m_1, \dots , m_{h+1}$, the term on the r.h.s. of (\ref{eq:ind}) associated with the indices $(r_1, \dots , r_{m_{h+1}})$, assuming that $r_{\alpha_1}, \dots r_{\alpha_k}$ close $k$ loops, with $1 \leq k \leq \min (m_1+ \dots + m_h, m_{h+1})$, while the other $m_{h+1}-k$ variables are fresh. Choose one of the $k$ loops, say the one linked with $r_{\alpha_1}$, denote by $s$ its length (by construction, $s\ge3$), and say it includes the edge $(1,2)$ (loops that do not involve the edge $(1,2)$ can be handled similarly). To bound the contribution of the integral associated with this choice of $(r_1, \dots , r_{m_{h+1}})$, we estimate $u_{h+1, r_{\alpha_j}}$ in $L^\infty$, for all $j=2,\dots , k$. After eliminating the dependence of the Bijl-Dingle-Jastrow function on their variables, we can then bound the remaining $m_1+\dots +m_{h+1} - (k-1) - (s -1)$ factors of $u$ that are not in the loop linked with $r_{\alpha_1}$ in $L^1$ (recall that we assumed the edge $(1,2)$ to be part of the loop; hence, the loop involves only $(s-1)$ factors of $u$). After appropriate renaming of the integration variables, this term can be estimated by 
\[  \begin{split} \Big|  \int V_{12} f_{12}^2 &\prod_{j_1=3}^{m_1+2} u_{1,j_1} \dots \prod_{j_h = m_1+\dots +m_{h-1} + 3}^{m_1 + \dots + m_h +2} u_{h, j_h} \prod_{j=1}^{m_{h+1}} u_{h+1, r_j}  \, \prod_{h+2 \leq i < j \leq N} f_{ij}^2 \Big| \\ \leq \; &C \| u \|_\infty^{k-1} \| u \|_1^{m_1+ \dots +m_{h+1} +2 - k - s} I_{N- (m_1 + \dots +m_{h+1} + 2 - k)} \\ &\hspace{4cm} \times \int V_{12} f_{12}^2 u_{2,3} u_{3,4} \dots u_{s-1,s} u_{1,s} dx_1 \dots dx_s\,. \end{split} \] 
With Lemma \ref{lm:lambdaell}, we can bound $\| u \|_\infty \leq 1$, $\| u \|_1 \leq C \frak{a} \ell^2$ and 
\[ \begin{split} \int V_{12} f_{12}^2 u_{2,3} &u_{3,4} \dots u_{s-1,s} u_{1,s} dx_1 \dots dx_s \\ &\leq C^s  \frac{\frak{a}^s |\Lambda |}{\ell^3}  \int \chi (|y_1 + \dots + y_{s-1} | \leq \ell)  \prod_{j=1}^{s-1} \frac{\chi_\ell (y_j)}{|y_j|} dy_1 \dots dy_{s-1} \\ &\leq C^s \frak{a}^s \ell^{2(s-1)-3}  |\Lambda| \,. \end{split} \]
Taking into account that $s \leq m_1 + \dots + m_{h+1} +2 - k$ and using (\ref{eq:INk}), this leads to 
\begin{equation*} \label{eq:loop2-fin} \begin{split} \Big|  \int V_{12} f_{12}^2 \prod_{j_1=3}^{m_1+2} u_{1,j_1} \dots & \prod_{j_h = m_1+\dots +m_{h-1} + 3}^{m_1 + \dots + m_h +2} u_{h, j_h} \prod_{j=1}^{m_{h+1}} u_{h+1, r_j}  \, \prod_{h+2 \leq i < j \leq N} f_{ij}^2 \Big| \\ \leq \; &C \frak{a}^2 \ell^{-1} (C \frak{a} \ell^2)^{m_1+ \dots +m_{h+1}- k} |\Lambda|  I_{N- (m_1 + \dots +m_{h+1} + 2 - k)} \\ \leq \; &C \rho \frak{a}^2 \ell^{-1} (C \frak{a} \ell^2/|\Lambda|)^{m_1+ \dots +m_{h+1}- k}  I_N / N\,.  
\end{split} \end{equation*}
 Thus, counting the number of terms on the r.h.s. of (\ref{eq:ind}) producing $k$ loops, we can estimate  
 \[ \begin{split} N &\cE_{\text{loops},h+1} \\ \leq \; &C \rho \frak{a}^2 \ell^{-1} \sum_{m_1=0}^M \binom{N-2}{m_1} \dots \sum_{m_h=0}^{M-m_1-\dots-m_{h-1}}  \binom{N-2 -\dots-m_{h-1}}{m_h} \chi(m_1+\dots+m_h \ge h-1)   \\ &\times \sum_{m_{h+1} = 0}^{M-m_1-\dots - m_h} \sum_{k=1}^{\min (m_{h+1}, m_1 + \dots + m_h)}  \binom{m_1 + \dots + m_h}{k} \binom{N-2 -m_1-\dots - m_h}{m_{h+1} - k} \\ &\hspace{9cm} \times  (C \frak{a} \ell^2/|\Lambda|)^{m_1+ \dots +m_{h+1}- k} I_N \\ \leq \; &C \rho \frak{a}^2 \ell^{-1} \sum_{m_1=0}^M \dots \sum_{m_h=0}^{M-m_1-\dots-m_{h-1}} \chi(m_1+\dots+m_h \ge h-1) \sum_{k=1}^{m_1 + \dots + m_h}   \binom{m_1 + \dots + m_h}{k} \\ &\times  \sum_{m_{h+1} = k}^{M-m_1-\dots - m_h}  \frac{N^{m_1+\dots +m_{h+1}-k}}{m_1! \dots m_h! (m_{h+1} -k)!}    (C \frak{a} \ell^2/|\Lambda|)^{m_1+ \dots +m_{h+1}- k} I_N\,.
 \end{split} \] 
Switching variables $m_{h+1} \to m_{h+1} - k$, we find 
\[ \begin{split} N &\cE_{\text{loops},h+1}  \\ \leq \; &C \rho \frak{a}^2 \ell^{-1} \sum_{m_1=0}^M \dots \sum_{m_h=0}^{M-m_1-\dots-m_{h-1}} \chi(m_1+\dots+m_h \ge h-1) \sum_{k=1}^{m_1+\dots + m_h}  \binom{m_1 + \dots + m_h}{k}  \\ &\times \sum_{m_{h+1} = 0}^{M-m_1-\dots - m_h-k}  \frac{1}{m_1! \dots m_h! m_{h+1}!}   (C \rho \frak{a} \ell^2)^{m_1+ \dots +m_{h+1}} I_N \,.
 \end{split} \] 
Next, we bound the sum over $m_{h+1}$ by $\exp (C\rho \frak{a} \ell^2) \leq C$ and subsequently the sum over $k$ by $2^{m_1 + \dots + m_h}$. Changing the value of the constant $C$, we arrive at
\[ \begin{split} N \cE_{\text{loops},h+1}  \leq \; &C \rho \frak{a}^2 \ell^{-1} \sum_{m_1=0}^M \dots \sum_{m_h=0}^{M-m_1-\dots-m_{h-1}}   \frac{ \chi(m_1+\dots+m_h \ge h-1) }{m_1! \dots m_h!}  (C \rho \frak{a} \ell^2)^{m_1+ \dots +m_{h}} I_N  \\ \leq \; &C  \rho \frak{a}^2 \ell^{-1}  (C \rho \frak{a} \ell^2)^{h-1} I_N \,.
 \end{split} \] 
Inserting in (\ref{eq:ind-1}) and then plugging the resulting bound in (\ref{eq:2body-indu}), we obtain (\ref{eq:2body-indu}), with $h$ replaced by $h+1$, completing the proof of the induction step. This proves the validity of (\ref{eq:2body-indu}), for all $h \in \bN$, $h \geq 2$. 

Choosing now $h=M$ in (\ref{eq:2body-indu}), we conclude that 
\begin{equation}\label{eq:M-iter} \begin{split}  \int V_{12} \prod_{1 \leq i < j \leq N} f_{ij}^2 \leq \; &|\Lambda| \Big[ 2\lambda_\ell \int \chi_\ell (x)  f_\ell^2 (x) dx \Big] \Big[ I_{N-2} + \sum_{k=3}^M \alpha_k I_{N-k} \| u \|_1^{k-2}  \Big] \\ &+ \int V_{12} f_{12}^2 \, \beta_M \prod_{M+1 \leq i < j \leq N} f_{ij}^2 + C \rho \frak{a}^2 \ell^{-1} \sum_{j=2}^M (C \rho \frak{a} \ell^2)^{j-2} I_N / N \, . 
 \end{split} \end{equation} 
The integral containing $\beta_M$ cannot be computed explicitly (some of the variables are still entangled with the Bijl-Dingle-Jastrow factor). With the definition of $\beta_M$, and using the bound $\| u \|_1 \leq C \frak{a} \ell^2$, following from Lemma \ref{lm:lambdaell}, we can estimate its absolute value by 
\[ \begin{split} 
\Big|  \int V_{12} f_{12}^2 \beta_M \prod_{M+1 \leq i < j \leq N} f_{ij}^2 \Big| \leq C &\; \lambda_\ell \ell^3|\Lambda| \sum_{m_1=0}^M \dots \sum_{m_M=0}^{M-m_1- \dots - m_{M-1}}  \chi (m_1 + \dots +m_M \geq M-1)  \\ &\times \frac{N^{m_1+\dots + m_M}}{m_1! \dots m_M!} \| u \|_1^{m_1 + \dots +m_M}  I_{N-(m_1+ \dots + m_M+2)}\,.  \end{split} \]
Taking into account the range of $m_1, \dots , m_M$, we decompose $\chi (m_1 + \dots +m_M \geq M-1) = \chi (m_1 + \dots + m_M = M-1) + \chi (m_1 + \dots + m_M = M)$. We find 
\begin{equation}\label{eq:Vbeta}  \begin{split} 
\Big|  \int V_{12} f_{12}^2 \beta_M \prod_{M+1 \leq i < j \leq N} f_{ij}^2 \Big| \leq \; &C \lambda_\ell \ell^3|\Lambda| N^{M-1} \| u \|_1^{M-1} I_{N-(M+1)} e^M\\&\hspace{4cm} + C  \lambda_\ell \ell^3|\Lambda| N^{M} \| u \|_1^{M} I_{N-(M+2)} e^M \\ \leq \; &C \rho \frak{a} \left[ (C \rho \frak{a} \ell^2 )^{M-1} +(C \rho \frak{a} \ell^2 )^M \right]  I_N/ N \leq C  \rho \frak{a} (C \rho \frak{a} \ell^2 )^{M-1} I_N / N\,.  \end{split} \end{equation} 
We conclude that 
 \begin{equation}\label{eq:V12-final} \begin{split} 
 \int V_{12} \prod_{1 \leq i < j \leq N} f_{ij}^2 \leq \; &|\Lambda| \Big[ 2 \lambda_\ell \int \chi_\ell (x) f_\ell^2 (x) dx \Big] \Big[ I_{N-2} + \sum_{k=3}^M \alpha_k I_{N-k} \| u \|^{k-2}  \Big] \\ &+C \rho \frak{a} (C \rho \frak{a} \ell^2)^{M-1} \, I_N / N + C \rho \frak{a}^2 \ell^{-1} \sum_{j=2}^M (C \rho \frak{a} \ell^2)^{j-2} I_N / N\,.
  \end{split} \end{equation} 

Similarly, we can bound the denominator on the l.h.s. of \eqref{eq:prop1} by 
\begin{equation} \label{eq:den-fin} \begin{split} I_N =  \int \prod_{1 \leq i < j \leq N} f_{ij}^2  \geq \; &|\Lambda| \Big[ \int f_\ell^2 (x) dx \Big] \Big[ I_{N-2} + \sum_{k=3}^{M} \alpha_k I_{N-k}  \| u \|_1^{k-2}  \Big] \\ &- C (C \rho \frak{a} \ell^2)^{M-1} \, I_N  - C \frak{a} \ell^{-1} \sum_{j=2}^M (C \rho \frak{a} \ell^2)^{j-2} I_N \,. \end{split} \end{equation} 
To prove this estimate, we proceed as in the derivation of (\ref{eq:V12-final}), replacing $V$ with $1$. Since we need here a lower rather than an upper bound, we replace $M$ with the odd integer $M+1$. This guarantees that whenever we expand part of the Bijl-Dingle-Jastrow function, we always stop the expansion with a negative contribution. As we did in the proof of (\ref{eq:M-iter}), we iterate $h=M$ times (despite the fact that we now cut expansions of the Bijl-Dingle-Jastrow function at order $M+1$, rather than $M$). Proceeding as in (\ref{eq:Vbeta}) to bound terms in which the integral cannot be performed explicitly, we arrive at (\ref{eq:den-fin}), with the coefficients $\alpha_k$ replaced by 
\[ \begin{split}  \wt{\alpha}_k = &\sum_{m_1=0}^{M+1} (-1)^{m_1} {N-2 \choose m_1} \dots \hspace{-.2cm}\sum_{m_{k-1} = 0}^{M+1-m_1-\dots - m_{k-2}} (-1)^{m_{k-1}} {N-2-m_1-\dots -m_{k-2} \choose m_{k-1}} \\ &\hspace{1cm} \times \Big[ \prod_{j=2}^{k-2} \chi (m_1 + \dots + m_j \geq j-1)  \Big] \, \chi (m_1 + \dots + m_{k-1} = k-2) \end{split} \]
for $k=1,\dots , M$. It is however easy to check that, due to the characteristic function $\chi (m_1 + \dots + m_{k-1} = k-2)$, the value of $\wt{\alpha}_k$ does not change if, on the r.h.s., we replace $M+1$ by $M$; in other words, $\wt{\alpha}_k = \alpha_k$, which leads to (\ref{eq:den-fin}). 

From (\ref{eq:den-fin}), we obtain (recall from (\ref{eq:ell}) that $\ell = c (\rho \frak{a})^{-1/2}$ so that $\frak{a} / \ell \leq C (\rho \frak{a}^3)^{1/2} \ll 1$) 
\[ I_N \geq |\Lambda| \Big[ \int f_\ell^2 (x) dx \Big] \Big[ I_{N-2} + \sum_{k=3}^{M} \alpha_k I_{N-k}  \| u \|_1^{k-2}  \Big] \Big[ 1 - C (C\rho \frak{a} \ell^2)^{M-1} - C \frak{a} \ell^{-1} \sum_{j=2}^M (C \rho \frak{a} \ell^2)^{j-2} \Big] \]
Combining with (\ref{eq:V12-final}), we arrive at 
\[ \begin{split} 
\frac{N}{2} \frac{\int V_{12} \prod_{i<j}^N f_{ij}^2}{I_N} \leq \; &N \frac{ \lambda_\ell \int \chi_\ell (x) f_\ell^2 (x) dx}{\int f_\ell^2 (x) dx}  \Big[ 1 + C (C \rho \frak{a} \ell^2)^{M-1} + C \frak{a} \ell^{-1} \sum_{j=2}^M (C \rho \frak{a} \ell^2)^{j-2} \Big] \\ &+ C \rho \frak{a} (C \rho \frak{a} \ell^2 )^{M-1} + C \rho \frak{a}^2 \ell^{-1} \sum_{j=2}^M (C \rho \frak{a} \ell^2)^{j-2} 
\,. \end{split} \]
Using Lemma \ref{lm:lambdaell}, we find 
\[ \lambda_\ell \int \chi_\ell (x) f_\ell^2 (x) dx \leq 4\pi \frak{a} \Big[ 1 + C \frac{\frak{a}}{\ell}
\Big]\,. \]
Since moreover $\int f_\ell^2 (x) dx \geq |\Lambda| - C \frak{a} \ell^2$, we conclude that 
\[ \begin{split} \frac{N}{2} \frac{\int V_{12} \prod_{i<j}^N f_{ij}^2}{\int \prod_{i<j}^N f_{ij}^2} \leq 4\pi \rho \frak{a} \Big[ 1 + C (C \rho \frak{a} \ell^2)^{M-1} + C \frak{a} \ell^{-1} \sum_{j=2}^M (C \rho \frak{a} \ell^2)^{j-2} \Big] 
\,. \end{split} \]
Choosing $\ell = c (\rho \frak{a})^{-1/2}$ as indicated in (\ref{eq:ell}), with $c > 0$ so small that, on the r.h.s. of the last equation, $C \rho \frak{a} \ell^2 \leq 1/2$, and choosing then the even number $M \geq 1 + \log_2 (\rho \frak{a}^3)^{-1/2}$, we obtain 
\[  \frac{N}{2} \frac{\int V_{12} \prod_{i<j}^N f_{ij}^2}{\int \prod_{i<j}^N f_{ij}^2} \leq 4\pi \rho \frak{a} \big[ 1 + C (\rho \frak{a}^3)^{1/2} \big] \, .\]
 
\section{Proof of Proposition \ref{prop:3body}} 

We proceed here similarly as in the proof of Prop. \ref{prop:2body}. For this reason, we will skip some of the details. As in (\ref{eq:ell}), we fix $\ell = c (\rho \frak{a})^{-1/2}$ for a sufficiently small constant $c >0$. 
 
Recalling the definition $u_{ij} = 1 - f_{ij}^2$ and the notation (\ref{eq:INk}), we set 
\[ \begin{split} \cE &= \frac{N^2}{I_N} \int \nabla f_{13}^2 \cdot \nabla f_{23}^2 \, f_{12}^2 \prod_{j \geq 4} f_{1j}^2 f_{2j}^2 f_{3j}^2 \prod_{4 \leq i < j \leq N} f_{ij}^2 \\ &=\frac{N^2}{I_N} \int \nabla u_{1,3} \cdot \nabla u_{2,3} \, f_{12}^2 \prod_{j \geq 4} f_{1j}^2 f_{2j}^2 f_{3j}^2 \prod_{4 \leq i < j \leq N} f_{ij}^2\,. \end{split} \]
With the bounds 
\begin{equation*}\label{eq:ubd}
0\le u(x) \le \frac{C \mathfrak{a}}{|x|} \chi_\ell (x) ,\qquad |\nabla u (x)|\le \frac{C\mathfrak{a}}{|x|^2} \chi_\ell (x) 
\end{equation*}
from Lemma \ref{lm:lambdaell} and with (\ref{eq:INm}) we find 
\[  N^2 \frac{I_{N-3}}{I_N} \int |\nabla u_{1,3}| |\nabla u_{2,3}|  \, u_{1,2}  \, dx_1 dx_2 dx_3  \leq C \rho^2  \frak{a}^3 \ell  \]
and therefore 
\begin{equation} \label{eq:starting_point_3body}
		\Big|\mathcal{E}-\frac{N^2}{I_N} \int \nabla u_{1,3} \cdot \nabla u_{2,3} \prod_{r=4}^N f^2_{1r}f^2_{2r} f^2_{3r} \prod_{4\le i <j\le N} f^2_{ij} \Big| \le C \rho \frak{a} (\rho \frak{a}^2 \ell ) \,.
	\end{equation}
Next, we expand the Bijl-Dingle-Jastrow factors, one variable after the other. Since here, in contrast with the proof of Prop. \ref{prop:2body}, the observable does not have a sign, when we stop an expansion we always have to estimate the error. We will use multiple times the inequality  
\begin{equation*} \label{eq:truncation}
	\begin{split}
		\Big| \prod_{j=r}^N f^2_{ij}& - \sum_{m=0}^k (-1)^{m} \sum_{r \le j_1 < \dots <j_{m} \le N} u_{i,j_1}\dots u_{i,j_{m}} \Big| \le  \sum_{r \le j_1<\dots< j_{k+1} \le N} \, u_{i,j_1}\dots u_{i,j_{k+1}} 
	\end{split}
\end{equation*}
which is valid for any $1\le i<r \le N$ and $k \ge 0$.
Applying this bound to (\ref{eq:starting_point_3body}), we find 
\begin{equation*}
	\begin{split}
		\Big|\mathcal{E}-&\frac{N^2}{I_N} \sum_{m_1=1}^M (-1)^{m_1} \binom{N-3}{m_1} \int \nabla u_{1,3} \cdot \nabla u_{2,3} \, u_{1,4}\dots u_{1,m_1+3}\prod_{r=4}^N f^2_{2r}f^2_{3r} \prod_{4\le i<j\le N}f^2_{ij} \Big|\\
		\le\;&C \frac{C^M N^{M+3}}{(M+1)! I_N} \int |\nabla u_{1,3}|\,|\nabla u_{2,3}| \, u_{1,4}\dots u_{1, M+4} \prod_{4 \le i<j\le N} f^2_{ij} + C\rho \frak{a} (\rho \frak{a}^2 \ell ) \\
		\le\;&C  \frac{C^M N^{M+3}}{(M+1)! I_N} \|\nabla u \|_1^2 \,\| u\|_1^{M+1} |\Lambda|  I_{N-(M+4)} + C \rho \frak{a} (\rho \frak{a}^2 \ell )  \\
		\le\;&  C \rho \mathfrak{a} (C \rho \frak{a} \ell^2)^{M+2} + C \rho \frak{a} (\rho \frak{a}^2 \ell ) 
	\end{split}
\end{equation*}
where in the last step we estimated $\| \nabla u \|_1 \leq C \frak{a} \ell$, $\| u \|_1 \leq C \frak{a} \ell^2$ and $I_N \geq 2^{-(M+4)} I_{N-(M+4)} |\Lambda|^{M+4}$ as follows from (\ref{eq:INm}). Notice that the sum on the l.h.s. starts from $m_1 = 1$, because the contribution with $m_1 = 0$ vanishes (since $\int \nabla u (x) dx = 0$). 

Let us now expand the $x_2$-dependence. We find
\begin{equation*} \label{eq:expansion_ell_2}
	\begin{split}
		\Big| \mathcal{E}-&\frac{N^2}{I_N} \sum_{m_1=1}^M (-1)^{m_1} \binom{N-3}{m_1} \sum_{m_2 = 1}^{M-m_1} (-1)^{m_2}  \sum_{4\le j_1<\dots< j_{m_2} \le N} \\ &\hspace{2cm} \times \int \nabla u_{1,3} \cdot \nabla u_{2,3}  \, u_{1,4} \dots u_{1, m_1+3} u_{2,j_1} \dots u_{2,j_{m_2}} \prod_{r=4}^N f^2_{3r} \prod_{4\le i<j\le N}f^2_{ij} \Big|	\\
		\le\;&C \frac{N^2}{I_N} \sum_{m_1=1}^M \frac{N^{m_1}}{m_1!} \sum_{4\le j_{1} <\dots < j_{M+1-m_{1} }\le N} \\ &\hspace{2cm} \times \int |\nabla u_{1,3}|\,|\nabla u_{2,3}|  \, u_{1,4} \dots u_{1,m_1+3} u_{2,j_1}\dots u_{2,j_{M+1 -m_1}} \prod_{4\le i<j\le N}f^2_{ij} \\
		&+ C \rho \mathfrak{a} (C\rho \frak{a} \ell^2)^{M+2} + C \rho \frak{a} (\rho \frak{a}^2 \ell )\,.
	\end{split}
\end{equation*}
Denoting by $0 \leq k \leq \min (m_1, M+1-m_1)$ the number of loops that are formed by the indices $j_1, \dots , j_{M+1-m_1}$, we can bound the first term on the r.h.s. of (\ref{eq:expansion_ell_2}) by 
\[ \begin{split} C &\frac{N^2}{I_N} I_{N-(M-k+4)} \sum_{m_1=1}^M \frac{N^{m_1}}{m_1!}  \sum_{k=0}^{\min (m_1, M+1-m_1)} \binom{m_1}{k} \binom{N-3-m_1}{M+1-m_1 - k} \\ & \times \int |\nabla u_{1,3}| |\nabla u_{2,3}|  \Big[ \prod_{j=4}^{k+3} u_{1,j} u_{2,j} \Big]  \, u_{1,k+4} \dots u_{1, m_1+3} u_{2, m_1+4} \dots u_{2, M-k+4} \, dx_1 \dots dx_{M-k+4} \\ \leq \; & C \sum_{k=0}^{(M+1)/2} \frac{1}{k!} \sum_{m_1 = k}^{M+1-k} \frac{1}{(m_1-k)! (M+1-m_1- k)!} 
 \rho^{M+3-k} (\frak{a} \ell^2)^{M+1-2k}  \frak{a}^{2k+2} \\ &\hspace{2cm} \times  \int \frac{\chi_\ell (x_1)}{|x_1|^2} \frac{\chi_\ell (x_2)}{|x_2|^2} \prod_{j=1}^k \frac{\chi_\ell (y_j)}{|y_j|} \frac{\chi_\ell (y_j + x_1 + x_2)}{|y_j + x_1 + x_2|} dx_1 dx_2 dy_1 \dots dy_k \\ \leq \; & C  \rho \frak{a} \sum_{k=0}^{(M+1)/2}  \frac{1}{k!} \sum_{m_1 = 0}^{M+1-2k} \frac{1}{m_1! (M+1-2k-m_1)!}  (C \rho \frak{a} \ell^2)^{M+2-2k} (C \rho \frak{a}^2 \ell)^k \\ \leq \; &C \rho \frak{a} \sum_{k=0}^{(M+1)/2} \frac{1}{k!} \frac{1}{(M+1-2k)!} (C \rho \frak{a} \ell^2)^{M+2-2k} (C \rho \frak{a}^2 \ell)^k \leq C\rho \frak{a} (C \rho \frak{a} \ell^2)^{M+2} + C \rho \frak{a} (\rho \frak{a}^2 \ell) \end{split} \] 
where, in the last step, we distinguish the cases $k=0$ and $k > 0$ (and we used the smallness of $\rho \frak{a} \ell^2$, resulting from (\ref{eq:ell})). Proceeding similarly, we can also bound the contribution of terms containing loops on the l.h.s. of (\ref{eq:expansion_ell_2}). We arrive at 
\[ \begin{split} \Big|  \mathcal{E} - &\frac{N^2}{I_N} \sum_{m_1=1}^M (-1)^{m_1} \binom{N-3}{m_1} \sum_{m_2 = 1}^{M-m_1} (-1)^{m_2} \binom{N-3-m_1}{m_2} \\ &\hspace{2cm} \times  \int \nabla u_{1,3} \cdot \nabla u_{2,3}  \, u_{1,4} \dots u_{1, m_1+3} u_{2,m_1+4} \dots u_{2, m_1+m_2+3} \prod_{r=4}^N f^2_{3r} \prod_{4\le i<j\le N}f^2_{ij} \Big|	\\
		\le\;&C \rho \mathfrak{a} (C \rho \frak{a} \ell^2)^{M+2} + C \rho \frak{a} (\rho \frak{a}^2 \ell )\,.
	\end{split}\]
Proceeding inductively (similarly as in the proof of Prop. \ref{prop:2body}), we find, after $M$ iterations,   
\[ \begin{split} \Big|  \mathcal{E}- \frac{N^2}{I_N} &\sum_{m_1=1}^M \sum_{m_2=1}^{M-m_1} \dots \sum_{m_M = 0}^{M-m_1 - m_2-\dots -m_{M-1}} (-1)^{m_1+\dots +m_M}\\ &\hspace{1cm} \times  \binom{N-3}{m_1} \dots \binom{N-3-m_1-\dots - m_{M-1}}{m_M}  \, \Big[ \prod_{j=5}^M \chi (m_1+ \dots + m_j \geq j-2) \Big]  \\ &\hspace{1cm} \times \int \nabla u_{1,3} \cdot \nabla u_{2,3}  \, \prod_{j_1=4}^{m_1+3} u_{1,j_1} \dots \prod_{j_M =m_1+\dots +m_{M-1} +4}^{m_1+\dots + m_M +3} u_{M,j_M}  \prod_{M+1 \le i<j\le N}f^2_{ij} \Big|	\\
		\le\;&C \rho \mathfrak{a} (C \rho \frak{a} \ell^2)^{M+2} + C \rho \frak{a} (\rho \frak{a}^2 \ell ) \sum_{j=2}^M (C \rho \frak{a} \ell^2)^{j-2} \,.
	\end{split}\]
The cutoffs $\chi (m_1 + \dots + m_j \geq j-2)$ make sure that, in all summands, the observable is still entangled with the Bijl-Dingle-Jastrow function. After removing contributions with loops (so that only trees are left), the cutoffs can be inserted for free, because $\int \nabla u (x) dx = 0$. 

Finally, estimating the absolute value of the sum on the l.h.s. of last equation by 
\[ \begin{split} 
C \frac{N^2}{I_N}  |\Lambda|  &\sum_{m_1, \dots , m_M = 0}^M  \frac{N^{m_1 + \dots + m_M}}{m_1! \dots m_M!} 
\chi (M-2 \leq m_1 +\dots + m_M \leq M) \\ &\hspace{3cm} \times  \| \nabla u \|_1^2  \| u \|_1^{m_1+ \dots + m_M}   I_{N-(m_1+\dots + m_M +3)}   \\  \leq \; &C \rho \frak{a} \, (C \rho \frak{a} \ell^2)^{M-1}  \end{split} \]
we conclude that 
\[ |\cE| \leq C \rho \mathfrak{a} (C \rho \frak{a} \ell^2)^{M-1} + C \rho \frak{a} (\rho \frak{a}^2 \ell )\,. \]
Recalling our choice of $\ell = c (\rho \frak{a})^{-1/2}$, fixing $c > 0$ so small that $C \rho \frak{a} \ell^2 \leq 1/2$ and
subsequently choosing the integer $M > 1 + \log_2 (\rho \frak{a}^3)^{-1/2}$, we obtain that 
\[ |\cE| \leq C \rho \frak{a} (\rho \frak{a}^3)^{1/2} \]
for a sufficiently large constant $C >0$. This concludes the proof of Prop. \ref{prop:3body}.

\medskip

{\it Acknowledgment.} We thank Giuseppe Scola for pointing us out Ref.\,\cite{PT}, with the proof 
of the cancellation of reducible diagrams within a convergent cluster expansion scheme. A.\,G. gratefully acknowledges financial support of the European 
Research Council through the ERC CoG UniCoSM, grant agreement n.\,724939, and
of MIUR, through the PRIN 2017 project MaQuMA, PRIN201719VMAST01. G.\,B., S.\,C., A.\,G. and A.\,O. warmly acknowledge  the GNFM Gruppo Nazionale per la Fisica Matematica - INDAM. B.\,S. gratefully acknowledges partial support from the NCCR SwissMAP, from the Swiss National Science Foundation through the Grant ``Dynamical and energetic properties of Bose-Einstein condensates'' (200020B\_200874) and from the European Research Council through the ERC-AdG CLaQS, grant agreement n.\,834782.  

\medskip

{\it Data availability.} Data sharing not applicable to this article as no datasets were generated or analysed during the current study.


\begin{thebibliography}{55}

\def \bibskip{\\[-0.43cm]}

\bibitem{BCOPS}
G. Basti, S. Cenatiempo, A. Olgiati, G. Pasqualetti, B. Schlein. A second order upper bound for the ground state energy of a hard-sphere gas in the Gross-Pitaevskii regime.  {\it Commun. Math. Phys.} (2022).  \bibskip



\bibitem{BCS}
G. Basti. S. Cenatiempo, B. Schlein. A new second order upper bound for the ground state energy of dilute Bose gases. {\it Forum Math. Sigma} {\bf 9} (2021), no. e74. \bibskip


\bibitem{B}
A. Bijl. The lowest wave function of the symmetrical many particles system. {\it Physica} {\bf 7} (1940), no. 9, 869-886. \bibskip

\bibitem{BBCS1}
C. Boccato, C. Brennecke, S. Cenatiempo, B. Schlein. Complete Bose-Einstein condensation in the Gross-Pitaevskii regime. {\it Commun. Math. Phys.} {\bf 359} (2018), no. 3, 975--1026. \bibskip


\bibitem{BBCS3}
C. Boccato, C. Brennecke, S. Cenatiempo, B. Schlein. Optimal rate for Bose-Einstein condensation in the Gross-Pitaevskii regime. {\it Commun. Math. Phys} {\bf 376} (2020), 1311--1395 .    \bibskip

\bibitem{BBCS-acta}
C. Boccato, C. Brennecke, S. Cenatiempo, B. Schlein. Bogoliubov Theory in the Gross-Pitaevskii limit. {\it Acta Mathematica} {\bf 222} (2019), no. 2, 219--335.    \bibskip


\bibitem{BSS1}
 C. Brennecke, B. Schlein, S. Schraven. Bose-Einstein Condensation with Optimal Rate for Trapped Bosons in the Gross-Pitaevskii Regime.  {\it Math. Phys. Anal. Geom.} {\bf 25} (2022), no.\,12. \bibskip
 

\bibitem{BSS2}
 C. Brennecke, B. Schlein, S. Schraven. Bogoliubov Theory for Trapped Bosons in the Gross-Pitaevskii Regime.   {\it Ann. Henri Poincar\'e.} {\bf 23} (2022), 1583--1658.  \bibskip

 \bibitem{SimpleEq2} 
 E.\,A. Carlen, M. Holzmann, I. Jauslin, E.H. Lieb. Simplified approach to the repulsive Bose
gas from low to high densities and its numerical accuracy. {\it Physical Review A} {\bf103} (2021), no. 6, 053309; Erratum {\it Phys. Rev. A} {\bf 104} (2021), 049904. \bibskip

\bibitem{CCS1}
C. Caraci, S. Cenatiempo, B. Schlein. Bose-Einstein condensation for two dimensional bosons in the Gross-Pitaevskii regime. {\it J. Stat. Phys. 183} {\bf 39} (2021). \bibskip
 

\bibitem{CCS2}
C. Caraci, S. Cenatiempo, B. Schlein. The excitation spectrum of two dimensional Bose gases in the Gross-Pitaevskii regime. Preprint arXiv:2205.12218.  \bibskip




\bibitem{CJL}
E.A. Carlen, I. Jauslin, E.H. Lieb. Analysis of a simple equation for the ground state energy
of the Bose gas. {\it Pure and Applied Analysis} {\bf (2)} (2020), no.3, 659--684. \bibskip



\bibitem{D} 
R. Dingle. The zero-point energy of a system of particles. {\it The London, Edinburgh, and Dublin
Philosophical Magazine and Journal of Science} {\bf 40} (1949), no. 304, 573--578. \bibskip

\bibitem{Dy}
F.J. Dyson. Ground-State Energy of a Hard-Sphere Gas. {\it Phys. Rev.} {\bf 106} (1957), 20--26. \bibskip


\bibitem{ESY}
L. Erd\H os, B. Schlein, H.-T. Yau. Ground-state energy of a low-density Bose gas: a second order upper bound. {\it Phys. Rev. A} {\bf 78} (2008), 053627. \bibskip


\bibitem{FGHP}
M. Falconi, E. L. Giacomelli, C. Hainzl, M. Porta. The Dilute Fermi Gas via Bogoliubov Theory. { \it Ann. Henri Poincare} {\bf 22} (2021), 2283--2353. \bibskip

\bibitem{FGJMO}
S. Fournais, T. Girardot, L. Junge, L. Morin, M. Olivieri. The Ground State Energy of a Two-Dimensional Bose Gas. Preprint arXiv:2206.11100.  \bibskip


\bibitem{FS1}
S. Fournais, J.P. Solovej. The energy of dilute Bose gases. {\it Ann. Math.} {\bf 192} (2020), no. 3, 893--976.  \bibskip

\bibitem{FS2}
S. Fournais, J.P. Solovej. The energy of dilute Bose gases II: The general case.
{\it Invent. Math.} (2022). \bibskip

\bibitem{G}
E. L. Giacomelli. An optimal upper bound for the dilute Fermi gas in three dimensions
Preprint arXiv:2212.11832.  \bibskip


\bibitem{HST} C. Hainzl, B. Schlein, A. Triay. Bogoliubov theory in the Gross-Pitaevskii limit revisited. 
Accepted for publication on {\it Forum Math. Sigma}. \bibskip

\bibitem{J} R. Jastrow. Many-body problem with strong forces. {\it Phys. Rev.} {\bf 98} (1955), no. 5, 1479--1484. \bibskip

\bibitem{La} 
A.B. Lauritsen. Almost optimal upper bound for the ground state energy of a dilute Fermi gas via cluster expansion. Preprint arXiv: 2301.08005. \bibskip

\bibitem{LaS} A.B. Lauritsen, R. Seiringer. Ground state energy of the dilute spin-polarized Fermi gas: Upper bound via cluster expansion. Preprint arXiv:2301.04894. \bibskip

\bibitem{LHY}
T. D. Lee, K. Huang, and C. N. Yang, Eigenvalues and eigenfunctions of a Bose
system of hard spheres and its low-temperature properties. {\it Phys. Rev.} {\bf 106} (1957), 1135--1145. \bibskip


\bibitem{L63} 
E.H.~Lieb. Simplified Approach to the Ground-State Energy of an Imperfect Bose Gas, {\it Phys.
Rev.}, {\bf 130} (1963), no. 6, 2518--2528. \bibskip


\bibitem{LS}
E.~H.~Lieb and R.~Seiringer.
\newblock Proof of {B}ose-{E}instein condensation for dilute trapped gases.
\newblock {\it Phys. Rev. Lett.} \textbf{88} (2002), 170409. \bibskip

\bibitem{LS2}
E.~H.~Lieb and R.~Seiringer.
\newblock Derivation of the Gross-Pitaevskii equation for rotating Bose gases.
\newblock {\it Comm. Math. Phys. } \textbf{264} (2006), no. 2, 505-537.  \bibskip

\bibitem{LSS}
E.~H.~Lieb, R.~Seiringer, J.~P.~Solovej.
Ground-state energy of the low-density Fermi gas.
{\it Phys. Rev. A} {\bf 71} (2005), 053605.  \bibskip

\bibitem{LSY}
E.~H.~Lieb, R.~Seiringer, and J.~Yngvason.
\newblock Bosons in a trap: A rigorous derivation of the {G}ross-{P}itaevskii
  energy functional. \newblock {\it Phys. Rev. A} \textbf{61} (2000), 043602.  \bibskip


\bibitem{LSY2}
E.H. Lieb, R. Seiringer, J. Yngvason. A Rigorous Derivation of the
Gross-Pitaevskii Energy Functional for a Two-dimensional Bose Gas. {\it Commun. Math. Phys.} \textbf{224} (2001), no. 1, 17--31. \bibskip


\bibitem{LY} 
E.~H.~Lieb, J.~Yngvason. Ground State Energy of the low density Bose Gas. {\it Phys. Rev. Lett.} {\bf 80} (1998), 2504--2507.   \bibskip

\bibitem{NNRT} 
P.~T.~{Nam}, M. Napi{\'o}rkowski, J. Ricaud, A. Triay. Optimal rate of condensation for trapped bosons in the Gross--Pitaevskii regime. {\it Anal. PDE} {\bf 15} (2022), no. 6, 1585--1616.  \bibskip

\bibitem{NRS}
P.~T.~{Nam}, N. ~{Rougerie}, R.~Seiringer. Ground states of large bosonic systems: The Gross-Pitaevskii limit revisited. {\it Analysis and PDE.} {\bf 9} (2016), no. 2, 459--485. \bibskip

\bibitem{NT}
P. T. Nam, A. Triay. Bogoliubov excitation spectrum of trapped Bose gases in the
Gross-Pitaevskii regime. Preprint arXiv:2106.11949. \bibskip

\bibitem{PB}
V. R. Pandharipande, H. A. Bethe. Variational Method for Dense Systems.
{\it Phys. Rev. C} {\bf 7} (1973), 1312--1328. \bibskip

\bibitem{PT}
E. Pulvirenti, D. Tsagkarogiannis. Cluster Expansion in the Canonical Ensemble
{\it Commun. Math. Phys.} {\bf 316} (2012), 289--306. \bibskip

\bibitem{YY}
H.-T. Yau, J. Yin. The second order upper bound for the ground state energy of a Bose gas. {\it J. Stat. Phys.} {\bf 136}(3) (2009), 453--503.


\end{thebibliography}
\end{document}